\documentclass[aps, prb, reprint, superscriptaddress]{revtex4-1}
\usepackage{graphicx}
\usepackage{dcolumn}
\usepackage{amsmath}
\usepackage{amssymb}
\usepackage{color}
\usepackage{bm}
\usepackage{braket}
\usepackage{txfonts}
\usepackage[colorlinks=true, letterpaper=true, pdfstartview=FitV, linkcolor=blue, citecolor=blue, urlcolor=blue]{hyperref}

\newcommand{\RNum}[1]{\uppercase\expandafter{\romannumeral #1\relax}}
\newcommand{\N}{\mathcal{N}}
\newcommand{\Tr}{\mathrm{Tr}}
\renewcommand{\S}{\overline{S}}
\newcommand{\RH}{\mathrm{H}}
\newcommand{\RL}{\mathrm{L}}
\newcommand{\RI}{\mathrm{I}}
\newcommand{\RT}{\mathrm{T}}
\newcommand{\hdd}{\hat{d}^\dag}
\newcommand{\hd}{\hat{d}}
\newcommand{\bv}{\bm{v}}
\newcommand{\bq}{\bm{q}}
\newcommand{\ba}{\bm{\alpha}}

\begin{document}

\title{Lindbladian dynamics with loss of quantum jumps}

\author{Yu-Guo Liu}
\affiliation{Beijing National Laboratory for Condensed Matter Physics, Institute of Physics, Chinese Academy of Sciences, Beijing 100190, China}

\author{Shu Chen}
\email{schen@iphy.ac.cn}
\affiliation{Beijing National Laboratory for Condensed Matter Physics, Institute of Physics, Chinese Academy of Sciences, Beijing 100190, China}
\affiliation{School of Physical Sciences, University of Chinese Academy of Sciences, Beijing 100049, China}

\date{\today}

\begin{abstract}
The Lindblad master equation (LME) describing the Markovian dynamics of the quantum open system can be understood as the evolution of the effective non-Hermitian Hamiltonian balanced with random quantum jumps. Here we investigate the balance-breaking dynamics by partly eliminating jumps from postselection experiments. To describe this dynamics, a non-linear Lindblad master equation (NLME) is derived from quantum trajectory method. However, the NLME shows significant advantages in analytical analysis over quantum trajectory method. Using the NLME, we classify the dynamics into two classes. In the trivial class, the process of reducing jumps is completely equivalent to weakening the coupling from the environment. In contrast, the nontrivial class presents more complex dynamics. We study a prototypical model within this class and demonstrate the existence of the postselected skin effect whose steady state is characterized by the accumulation of particles on one side. The steady-state distribution can be fitted by a scale-invariant tanh function which is different from the uniform distribution of LME.  Furthermore, the NLME can give a reasonable framework for studying the interplay and competition between the non-Hermitian Hamiltonians and dissipative terms. We show this by capturing the characteristics of the trajectory-averaged entanglement entropy influenced by non-Hermitian skin effect and Zeno effect in the model of postselected skin effect.
\end{abstract}

\maketitle


\section{Introduction}
The Lindblad master equation (LME) is one of the most successful dynamics equations in the quantum open system~\cite{Lindblad1976,Gorini1976,Gardiner1985,Daniel2020}. Its success is not only attributed to the effective description of the quantum system weakly coupling to the environment but also stems from its concise mathematical structure with good properties such as trace-preserving, trace-preserving, and complete positivity. From the perspective of quantum trajectory theory~\cite{Dalibard1992,Dum1992,Gardiner1992,Klaus1993,Plenio1998,Daley2014}, the LME can be understood as an average of many quantum trajectories consisting of the evolution of non-Hermitian Hamiltonian (ENHH) and random quantum jumps. Due to the constraint by LME, ENHH and the quantum jumps are required to reach a balance in the average sense of all trajectories to keep some good properties like trace-preserving.

In recent years, non-Hermitian systems have attracted much attention from both theoretical and experimental studies~\cite{Yao2018,Kunst2018,CHLee2019,FSong2019,CHLiu2020,PHe2022,Haga2021,FYang2022,ZGong2018,NOkuma2023,HJiang2019,DWZhang2020,Hugo2021,CXG2021,CXG2023,Niu2023,WDing2023,WGou2020,LXiao2020,CWang2023,TELee2014,KYamamoto2019,ZXu2020,TLiu2020,KYang2021,Kawabata2023,KLi2023,SZLi2024,LZ2024,JLi2019,Naghiloo2019,YWu2019,WChen2021,Minganti2020,Lewalle2023}. It is often regarded as a method to obtain the ENHH by postselecting the quantum trajectories with no jumps~\cite{TELee2014,KYamamoto2019,ZXu2020,TLiu2020,KYang2021}. However, finding a pure non-Hermitian trajectory without jumps is difficult due to the probability tending to zero with increasing observation time and jump operators.  A more viable option to reflect non-Hermitian physics is only partly eliminating jumps, then the ENHH can obtain advantages in the competition with quantum jumps. Notably, it will break the balance of LME even if a few quantum jumps are discarded. This motivated us to study the balance-breaking dynamics of Lindblad master equations with the loss of quantum jumps.

In Ref. \cite{Minganti2020}, Minganti et al. propose postselection methods to partly discard jumps, which can be described by a linear hybrid-Liouvillian formalism. This formalism does not preserve the unit trace due to a lack of normalization in their processes. However, in many theoretical and experimental researches, normalization is required, for example, the populations of atoms or qubits are normalized in experiments of non-Hermitain systems~\cite{JLi2019,Naghiloo2019,YWu2019,WChen2021}, and the wavefunctions are continuously normalized in the theoretical studies of entanglement growth influenced by measurements~\cite{Li2018,Skinner2019,Li2019,Chan2019,Cao2019,Alberton2021,Turkeshi2021,Coppola2022,YNZhou2023,Wang2022,Feng2023} or non-Hermitian Hamiltonians~\cite{Kawabata2023,KLi2023,SZLi2024,LZ2024}. Usually, the non-linearity will arise from the normalization, as shown in the work of the generalized Lindblad master equation by Zhou~\cite{YNZhou2023} and the work of Lewalle et al. for a non-Hermitian single qubit~\cite{Lewalle2023}. Therefore what we concern here, different from the work of Minganti et al., are the general properties of a non-linear equation and special phenomena of the normalized postselection dynamics.


In this work, we use the quantum trajectory method to derive a nonlinear Lindblad master equation (NLME) to describe the dynamics of an open system lacking an arbitrary proportion of quantum jumps, which continuously interpolates between LME and ENHH. The NLME has greater advantages in analytical analysis than the quantum trajectory method itself, which allows us to obtain the fundamental properties of the postselection dynamics and gives us a chance to get the analytical solution for some models. Within the framework of NLME, we answer the question: how does the dynamics be influenced by discarding quantum jumps?  Basically, reducing jumps can slow down the losses of quantum coherence and purity in the sense of postselection. Further, we find the situations can be divided into two classes. For the trivial class, the process of reducing jumps is completely equivalent to weakening the coupling from the environment, where the NLME is reduced to LME. For the nontrivial class, loss of jumps might bring totally different relaxation process and steady state, compared to LME. Here we give a prototypical nontrivial model to showcase a new phenomenon, the postselected skin effect, where particles in the non-equilibrium steady state will accumulate on one side as long as an arbitrarily small proportion of quantum jumps is discarded by postselection.

Moreover, in the usual studies, the system Hamiltonian of LME is Hermitian. However, the NLME inherently contains the effective non-Hermitian Hamiltonian and pure dissipative terms, providing a reasonable framework for the study of non-Hermitian systems embedded in open environment. We show this in our model for the postselected skin effect, in which the trajectory-averaged entanglement entropy (TAEE) is influenced by the coexistence and competition between the skin effect from the non-Hermitian Hamiltonian and the Zeno effect from the pure measurements.

\section{Nonlinear Lindblad master equation}
\begin{figure}[htbp]\centering
\includegraphics[width=8cm]{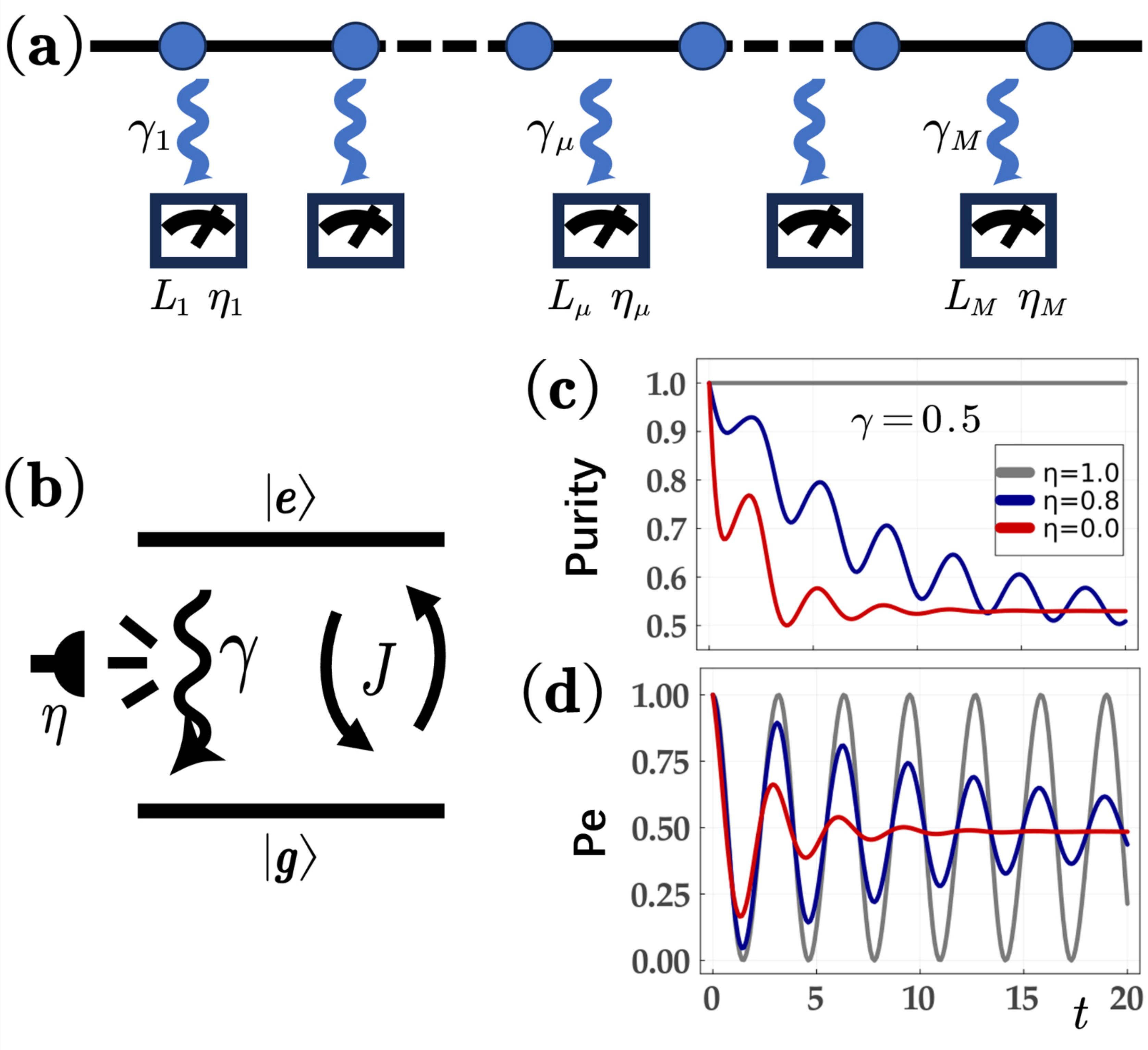}
\caption{ Illustration of the postselection experiments in  chain and two-level atom. (a) The system is coupled to the environment by jump operators $L_\mu$ with coupling strength $\gamma_\mu$. Every jump is monitored by a detector with efficiency $\eta_\mu$. Only the experimental data with no quantum jumps detected is preserved. (b) Postselection on a two-level atom with spontaneous emission. Emission photons are detected with efficiency $\eta$. Initialize the atom at $|e\rangle$. Observe the population of two levels. Discard the observed data when a photon is detected at the same time. After the postselection, the purity of the atomic density matrix $\Tr (\rho^2)$ and the probability of $|e\rangle$ are shown in (c) and (d). With increasing $\eta$ from $0$ (red line) to $0.8$ (blue line) and then to $1$ (grey line), the purity increases and Rabbi oscillation time becomes longer. In (c) and (d), $J=1$ and $\gamma=0.5$}
\label{fig::1}
\end{figure}

Considering a system with Hamiltonian $H$ coupling to the environment by $M$ jump operators $L_\mu$ with strength $\gamma_\mu$, its density matrix $\rho$ is governed by LME:
\begin{equation}\label{LME}
\frac{d\rho}{dt} =\mathcal{L}(\rho):= -i[H,\rho] + \sum_{\mu=1}^M \gamma_{\mu} \left( L_{\mu} \rho L_{\mu}^{\dag} - \frac{1}{2} \{ L_{\mu}^{\dag} L_{\mu}, \rho \} \right),
\end{equation}
where $\mathcal{L}$ is the Liouvillian superoperator and Planck constant $\hbar$ is set to unity throughout this paper. We monitor every jump by a $\eta_\mu$-efficiency detector, $\eta_\mu \in [0,1]$, as shown in Fig.~\ref{fig::1}(a). The core concern of the present work is how $\rho$ evolves after discarding the data along with detectors clicking. When $\eta_\mu=0$, the detector does not work, meaning all quantum trajectories are retained, and the dynamics is fully described by the LME. Conversely, for an ideal detector ($\eta_\mu=1$), every quantum jump is perfectly monitored, and all trajectories containing jumps are discarded, resulting in postselection of purely ENHH. For a finite efficiency detector, the quantum trajectories will be probabilistically discarded, so the postselected process is the Lindbladian dynamics with loss of part of quantum jumps.

The equation of motion is derived in the spirit of the quantum trajectory method.  At time $t$, the system is assumed in a pure state $|\phi (t)\rangle$. Postselecting the data after $\delta t$, the $L_\mu$ will make the system jump to the state $\N L_\mu |\phi (t)\rangle$ with the probability of $(1-\eta_\mu) \delta p_\mu$ or evolve to the state $\N exp(-\frac{1}{2}\gamma_\mu L^\dag_{\mu}L_{\mu}\delta t)|\phi (t)\rangle$, where $\N$ denotes normalization and $\delta p_\mu = \gamma_\mu \delta t \langle \phi(t)|L^\dag_{\mu}L_{\mu} |\phi(t) \rangle$. Then the state at time $t+\delta t$ is
\begin{equation}\label{tadt}
|\phi(t+\delta t)\rangle=e^{-iH\delta t} \N \prod_{\mu=1}^{M}\left\{ \begin{matrix} (1-\eta_\mu) \delta p_\mu : & L_\mu \\ 1-(1-\eta_\mu) \delta p_\mu : & e^{-\frac{1}{2}\gamma_\mu L^\dag_{\mu}L_{\mu}\delta t}\end{matrix}  \right\} |\phi(t)\rangle.
\end{equation}
The error from the non-commutativity among different jump operators and Hamiltonian is ignored with $\delta t \to 0$. By using
\begin{equation}
\frac{d}{dt} \rho = \lim_{\delta t \to 0} \frac{1}{\delta t} (|\phi(t+\delta t)\rangle \langle \phi(t+\delta t) | - |\phi(t)\rangle \langle \phi(t) |),
\end{equation}
we get the NLME (see Appendix ~\ref{AP:1} for details):
\begin{equation}\label{NLME}
\begin{split}
& \frac{d}{dt} \rho = \mathcal{L}_{\eta}(\rho):= -i[H,\rho] + \\
& \sum_{\mu=1}^M \gamma_{\mu} \left( -\frac{1}{2} \{ L_{\mu}^{\dag} L_{\mu}, \rho \} + (1-\eta_\mu) L_{\mu} \rho L_{\mu}^{\dag} + \eta_\mu  \langle L_{\mu}^{\dag} L_{\mu} \rangle \rho \right) ,
\end{split}
\end{equation}
where the nonlinear term $\langle L_{\mu}^{\dag} L_{\mu} \rangle \rho= \Tr(L_{\mu}^{\dag} L_{\mu} \rho)\, \rho$ comes from the normalization process. We note that the single-qubit version of this equation related with spontaneous emission also obtained in Ref.~\cite{Lewalle2023}.

From the perspective of postselection, it is natural to rewrite Eq.~(\ref{NLME}) as a combination of effective non-Hermitian Hamiltonian $H_{eff}$ and quantum jumps $\mathcal{L}_{QJ}$:
\begin{equation}\label{formA}
\frac{d}{dt} \rho = -iH_{eff}(\gamma)\rho + i\rho H^\dag_{eff}(\gamma) + \mathcal{L}_{QJ}(\rho) + \mathcal{L}_{\N}(\rho),
\end{equation}
where
\begin{subequations}
\begin{align}
& H_{eff}(\gamma)=H-\frac{i}{2}\sum_\mu \gamma_\mu L_{\mu}^{\dag} L_{\mu}, \\
& \mathcal{L}_{QJ}(\rho) = \sum_\mu (1-\eta_\mu)\gamma_\mu L_\mu \rho L_\mu^\dag, \\
& \mathcal{L}_{\N}(\rho) = \sum_\mu \eta_\mu \gamma_\mu  \langle L_{\mu}^{\dag} L_{\mu} \rangle \rho,
\end{align}
\end{subequations}
where $\mathcal{L}_{\N}$ is the normalization term. Alternatively, Eq.~(\ref{NLME}) can also be represented as a combination of effective non-Hermitian Hamiltonian and pure dissipative or pure measurements terms $\mathcal{L}_{M}$:
\begin{equation}\label{formB}
\frac{d}{dt} \rho = -iH_{eff}(\eta\gamma)\rho + i\rho H^\dag_{eff}(\eta\gamma) + \mathcal{L}_{M}(\rho) + \mathcal{L}_{\N}(\rho),
\end{equation}
where
\begin{subequations}
\begin{align}
& H_{eff}(\eta \gamma)=H-\frac{i}{2}\sum_\mu \eta_\mu \gamma_\mu L_{\mu}^{\dag} L_{\mu},\\
& \mathcal{L}_{M}(\rho)=\sum_\mu (1-\eta_\mu) \gamma_\mu \left( L_{\mu} \rho L_{\mu}^{\dag} - \frac{1}{2} \{ L_{\mu}^{\dag} L_{\mu}, \rho \}\right).
\end{align}
\end{subequations}

The simplest model for Eq.~(\ref{NLME}) is a two-level atom with monitoring its spontaneous emission, as shown in Fig.~\ref{fig::1}(b), where $J$ is the coupling strength and $\gamma$ is the spontaneous emission rate of the excited state $|e\rangle$. The detector of emission photons has a finite efficiency $\eta$. This setup is also discussed in Ref.~\cite{Lewalle2023}. In our  theoretical study, the observed data of atomic population is discarded when emission photons are detected simultaneously. With the remaining data, the evolution of atomic system is rebuilt by
\begin{equation}~\label{AtomE}
\frac{d}{dt} \rho = -i[H,\rho] +  \gamma \left( -\frac{1}{2} \{ L^{\dag} L, \rho \} + (1-\eta) L \rho L^{\dag} + \eta  \langle L^{\dag} L \rangle \rho \right),
\end{equation}
where $H=J(|e\rangle \langle g|+h.c.)$ and $L=|g\rangle \langle e|$. Starting from $|e\rangle$, the evolutions of the purity $\Tr (\rho^2)$ and excited state probability $P_e$ are shown in Fig.~\ref{fig::1}(c) and Fig.~\ref{fig::1}(d), where with increasing $\eta$, the purity becomes larger and Rabbi oscillation becomes longer. This reflects a basic fact that decreasing quantum jumps can slow down the loss rate of purity and quantum coherence. 

A vectorization treatment of Eq.~(\ref{AtomE}) is given in Appendix ~\ref{AP:2}. We further compare the results from this method with results from two kinds of quantum trajectory methods in Appendix ~\ref{AP:4} to confirm the validity of NLME.

\section{Properties and classification}
The NLME given by Eq.~(\ref{NLME}) provides a convenient method to study the postselection dynamics. Before going to the study of concrete systems, we discuss some specific features of the NLME. Firstly, we show some properties of NLME.

Property I: $\mathcal{L}_{\eta}(\rho)^\dag = \mathcal{L}_{\eta}(\rho^\dag)$. If the initial density matrix is Hermitian, i.e. $\rho_0^\dag=\rho_0$, the NLME is Hermiticity-preserving, i.e. $\rho(t)^\dag=\rho(t)$, where $\rho(t)=e^{\mathcal{L}_{\eta} t}\,\rho_0 $.

Property II: $\frac{d}{dt} \, \Tr (\rho)=\sum_\mu \eta_\mu \gamma_\mu (\Tr(\rho)-1)\langle L_{\mu}^{\dag} L_{\mu} \rangle$. If $\Tr(\rho_0)=1$, the NLME is trace-preserving, i.e. $\Tr (\rho(t))=1$.

Property III: No conservation under strong symmetry. For the LME in Eq.~(\ref{LME}), the strong symmetry condition
\begin{equation}
[U,H]=[U,L_\mu]=0,\  for\  all\  \mu,
\end{equation}
can guarantee the conservation of the observation $O$, where $U=e^{i\theta O}$ is the continuous symmetry with $\theta \in \mathbb{R}$~\cite{Prosen2012,Albert2014}. However, for the NLME, the strong symmetry condition leads to
\begin{equation}
\frac{d}{dt}\Tr(O\rho)=\sum_\mu \eta_\mu \gamma_\mu (\Tr(O\rho)-1) \langle L_{\mu}^{\dag} L_{\mu} \rangle,
\end{equation}
where a non-zero $\eta_\mu$ can break the conservation of $O$.

Properties I and II unveil that the NLME is Hermiticity-preserving and trace-preserving for a real physics process. Property III indicates that the nonlinearity breaks the conservation law.

Classification: According to whether it can be reduced to LME, the NLME can be divided into trivial and nontrivial classes. When $L_\mu^\dag L_\mu \rho(t)=\lambda_\mu \rho(t)$ holds for all $\mu$, the NLME becomes
\begin{equation}
\frac{d\rho}{dt} = -i[H,\rho] + \sum_{\mu=1}^M (1-\eta_\mu)\gamma_{\mu} \left( L_{\mu} \rho L_{\mu}^{\dag} - \frac{1}{2} \{ L_{\mu}^{\dag} L_{\mu}, \rho \} \right),
\end{equation}
which is just the LME in Eq.~(\ref{LME}) with $\gamma_\mu \to \gamma_\mu^{eff}=(1-\eta_\mu)\gamma_{\mu}$. We call this case as a trivial class where the postselection can be regarded as a weakening of the coupling from the environment. There is a sufficient condition for the trivial class:
\begin{equation}\label{TivalC}
L_{\mu}^{\dag} L_{\mu} \rho_0=\lambda_\mu \rho_0, \ [L_{\mu}^{\dag} L_{\mu},H]=[L_{\mu}^{\dag} L_{\mu},L_k]=0\  for\ all\ \mu,\, k,
\end{equation}
where $\rho_0$ is the initial state. Systems decaying by Pauli operators belong to the trivial class, because $L_{\mu}^{\dag} L_{\mu}$ is an identity matrix. For a translational invariant system, the condition~(\ref{TivalC}) expands to
\begin{equation}\label{TivalTIC}
\Gamma \rho_0=\lambda \rho_0, \ [\Gamma,H]=[\Gamma,L_k]=0\  for\ all\ k,
\end{equation}
where $\Gamma=\sum_\mu L_{\mu}^{\dag} L_{\mu}$.

\section{An example in the trivial class}
Considering a homogeneous fermion chain in periodic boundary condition under $\eta$-efficiency continuous measurements of particle number in every site. The postselection dynamics is described by

\begin{equation}\label{MPN}
\frac{d\rho}{dt}  = -i[H,\rho] + \gamma \sum_l \left(  -\frac{1}{2}\{ \hat{n}_l^2, \rho \} + (1-\eta) \hat{n}_l \rho \hat{n}_l  + \eta  <\hat{n}_l^2>\rho   \right),
\end{equation}
where $H=\sum_{mn}\mathcal{H}_{mn}a_m^\dag a_n $ is a quadratic system Hamiltonian, $\gamma$ is the measurement strength of local particle-number operator $\hat{n}_l=a_l^\dag a_l$.

The correlation Matrix $G$ is defined by $G_{ij} := <a^\dag_i a_j>=\Tr(a^\dag_i a_j \rho)$. By Eq.~(\ref{MPN}), we have
\begin{equation}
\begin{split}
\frac{d}{dt}G_{ij} = & -i(\mathcal{H}^\RT G - G \mathcal{H}^\RT )_{ij} -(1-\eta)\gamma(1-\delta_{ij}) G_{ij} \\
                     &+ \eta \gamma \left(<\hat{N}>G_{ij} - <a^\dag_i a_j  \hat{N}>\right),
\end{split}
\end{equation}
and
\begin{equation}
\frac{d}{dt}<\hat{N}>=\frac{d}{dt}\sum_l G_{ll}=\eta \gamma \left( <\hat{N}>^2 - <\hat{N}^2> \right),
\end{equation}
where $\hat{N}=\sum_l \hat{n}_l$ is the total particle number operator. When $ \eta \gamma \ne 0$, the conservation of total particle number requires the fluctuation $<\hat{N}>^2 - <\hat{N}^2>$ is $0$, namely $\rho$ is the eigen matrix of $\hat{N}$.

We have $[\hat{N},H]=[\hat{N},\hat{n}_l]=0$ and when the initial state is the eigenstate of $\hat{N}$, the trivial condition Eq.~(\ref{TivalTIC}) is satisfied. Then the Eq.~(\ref{MPN}) is reduced to
\begin{equation}
\frac{d\rho}{dt} =-i[H,\rho] + (1-\eta) \gamma \left(  -\frac{1}{2}\{ \hat{N}, \rho \} + \sum_l \Big(  \hat{n}_l \rho \hat{n}_l  \Big) \right).
\end{equation}
The above equation is just the LME with the substitution $\gamma \to (1-\eta)\gamma$. It illustrates that, in the trivial class, increasing postselcted strength $\eta$ in NLME is completely equivalent to decreasing the measurement strength $\gamma$ in LME.

The above analysis  shows the powerful analytical ability of NLME. Especially for a single site case (the length of the chain is 1 in Eq.~(\ref{MPN})), the equation of particle number becomes
\begin{equation}
\frac{d}{dt}   \Tr(\hat{n}\rho)=-\gamma \eta  \Tr(\hat{n}\rho) + \gamma \eta\ \Tr(\hat{n}\rho)^2.
\end{equation}
Then the particle number at time $t$ has the analytic expression:
\begin{equation}
n(t)=\frac{1}{1+(\frac{1}{n_0}-1)e^{\eta \gamma t}},
\end{equation}
where $n_0$ is the average particle number of the initial state.

\section{Postselected skin effect}
\begin{figure}[htbp]
\includegraphics[width=8.5cm]{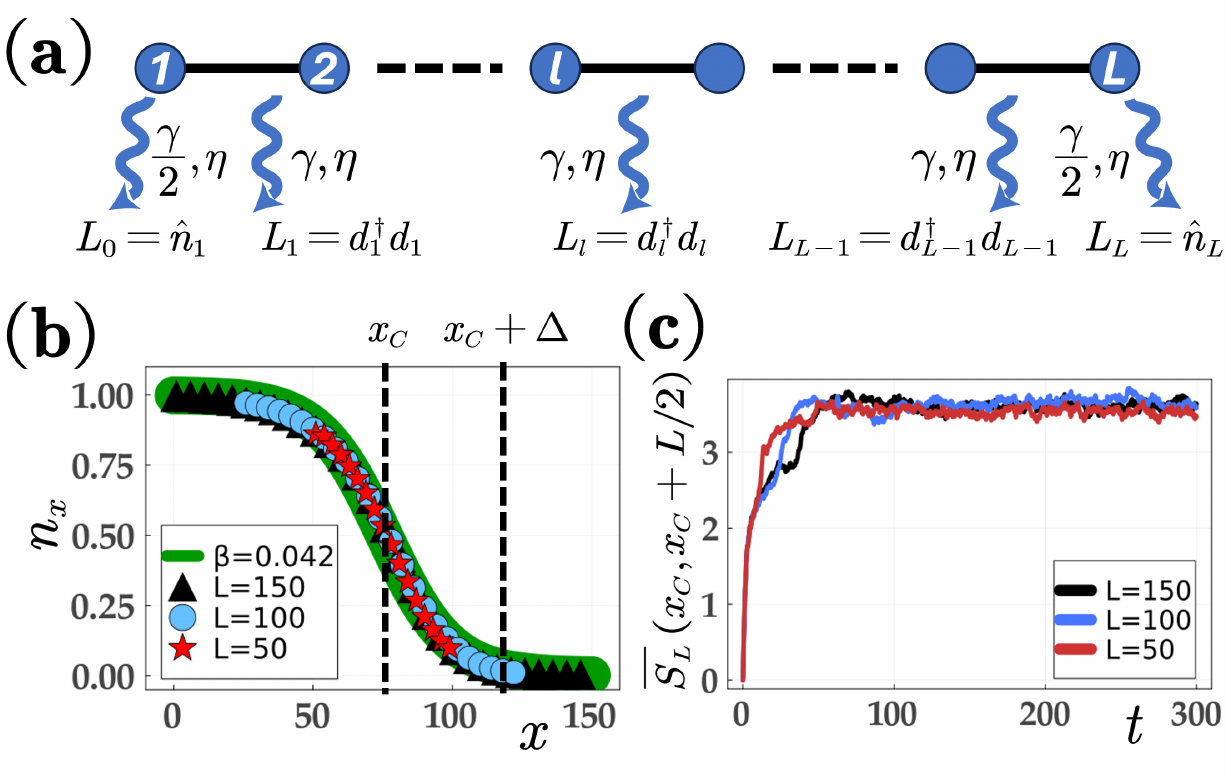}
\caption{ Postselected skin effect. (a) Illustration of the model with postselected skin effect. (b) Steady-state distributions of the half-filled chain with 50 (red star), 100 (blue circle), and 150 (black triangle) sites. The green line is the fitting function Eq.~(\ref{TanhX}) with $\beta=0.04$. (c) Evolution of half-chain entanglement entropy defined by Eq.~(\ref{TAEE}) in the half-filled chain with 50 (red line), 100 (blue line), and 150 (black line) sites. In (b) and (c), $\gamma=0.4$, $\eta=0.6$ and the initial state is $|1010\cdots\rangle$. The results are an average of $60$ quantum trajectories by the Gaussian state method and the Monte Carlo wave-function method (see Appendix~\ref{AP:3}) with time step $\delta t= 0.005$.}
\label{fig::2}
\end{figure}

Now we study a model in the nontrivial class, which exhibits skin effect induced by postselection. Considering a $L$-site free fermion chain under open boundary condition monitored by $L+1$ detectors with unified efficiency $\eta$ as shown in Fig.~\ref{fig::2}(a), the corresponding NLME is given by
\begin{equation}\label{PSE}
\begin{split}
&\frac{d\rho}{dt} = -i[H,\rho] +\\
&\sum_{l=0}^{L} \gamma_l \left( - \frac{1}{2} \{ L_l^{\dag} L_l, \rho \} + (1-\eta) L_l \rho L_l^{\dag} + \eta \langle L_l^\dag L_l \rangle \rho \right),\\
\end{split}
\end{equation}
where $H=J\sum_{l=1}^{L-1} a_l^\dag a_{l+1} + h.c.$, $\gamma_l=\gamma$, $L_l=d_l^\dag d_l$  with $d_l=(a_l + ia_{l+1})/\sqrt{2}$ for $l=1\sim L-1$,  $\gamma_0=\gamma_L=\gamma/2$, $L_0=\hat{n}_1$ and $L_L=\hat{n}_L$. For simplicity, we set $\gamma \le J=1$. We note that the jump operator $L_l$ originates from the Ref.~\cite{Wang2022}, where a biased steady state is induced through the combination between $L_l$ and some feedback unitary operators. In this work, we demonstrate that a biased steady state can also be achieved by replacing the feedback unitary operators with postselection.

\begin{figure*}[htbp]\centering
\includegraphics[width=17cm]{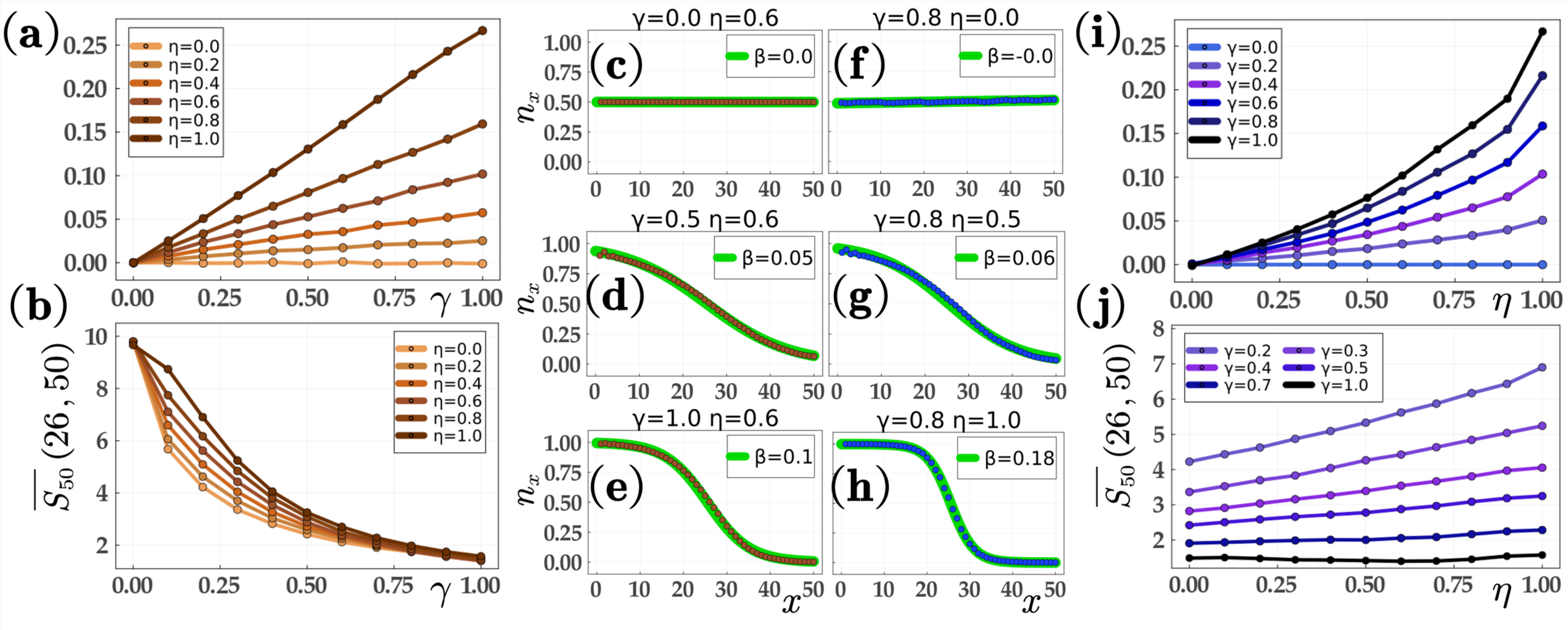}
\caption{ Steady-state characteristics of postselected skin effect. (a) The fitting parameter $\beta$ and (b) the half-chain entanglement entropy $\S_L(x_C,x_C+L/2)$ of the steady state as a function of $\gamma$ with different $\eta$, where $\beta$ has linear relationship with $\gamma$ implying a scale invariance of steady state in terms of $1/\gamma$. (c) $\sim$ (h) Steady-state distribution with different $\gamma$ and $\eta$ (points from numerical results and green lines from function fitting). (i) $\beta$ and (j) $\S_L(x_C,x_C+L/2)$ of the steady state as a function of $\eta$ with different $\gamma$, where the entropy lines show different monotonicity of $\eta$ as a result from the combining of Zeno effect and skin effect. In all subfigures, results are from an average of $60$ quantum trajectories in a half-filled $50$-site chain by Monte Carlo wave-function method with time step $\delta t= 0.005$.}
\label{fig::3}
\end{figure*}

When $\eta=0$, the evolution equation reduces to LME. The Hermiticity of $L_l$ ensures the identity matrix is a steady-state solution of Eq.~(\ref{PSE}), which means the steady state exhibits a uniform distribution on the lattice, as shown in Fig.~\ref{fig::3}(f). However, when $\eta \ne 0$, the particles will accumulate on one side, as shown in Fig.~\ref{fig::2}(b). The steady-state population in a half-filled chain is well fitted by a tanh function:
\begin{equation}\label{TanhX}
n(x)=-\frac{1}{2}\tanh\left(\beta \,(x-\frac{L+1}{2}) \right)+0.5,
\end{equation}
where $n(x)$ is the particle number on the site $x$. The fitting parameter $\beta$ is determined by $\gamma$ and $\eta$, and it is almost unchanged for different lattice sizes. In Fig.~\ref{fig::2}(b), the steady states of different sizes are fitted with the same $\beta$ (green line).  Since the skin steady state is induced by a postselected  processes, we call this phenomenon as the postselected skin effect.

To explain the postselected skin effect, we rewrite Eq.~(\ref{PSE}) as Eq.~(\ref{formB}), with
\begin{subequations}
\begin{align}
& H_{eff}(\eta \gamma) = \sum_{l=1}^{L-1} \left( (J-\eta \gamma/4)a_{l+1}^\dag a_l + (J+ \eta \gamma/4) a_l^\dag a_{l+1}    \right), \label{NHT}\\
& \mathcal{L}_{M}(\rho) = \sum_{l=0}^L (1-\eta)\gamma (1-\frac{\delta_{l0}+\delta_{lL}}{2}) \left(L_l \rho L^\dag_l-\frac{1}{2}\{L_l^\dag L_l, \rho\}\right), \label{MeaT}\\
& \mathcal{L}_{\N}(\rho) = -\frac{1}{2}\eta \gamma \left \langle \sum_{l=1}^{L-1} j_l \right \rangle \rho,
\end{align}
\end{subequations}
where $j_l = -i(a^\dag_{l+1}a_l - a^\dag_l a_{l+1}) $. Since the pure measurement term $\mathcal{L}_{M}$ and the normalization term $\mathcal{L}_{\N}$ do not offer a bias force on particles, the non-Hermitian $H_{eff}$ as a Hatano-Nelson model~\cite{Hatano1996,Hatano1997} will drive particles gathering on the left side, as long as $\eta \gamma \ne 0$.

The value of $\beta$ reflects the strength of postselected skin effect. There are two mechanisms to increase $\beta$. One is increasing $\gamma$, which transfers the dynamics from a non-dissipative free system to NLME.  Populations of steady states with different $\gamma$ are shown in Figs.~\ref{fig::3} (c) $\sim$ (e). The other one is increasing $\eta$, transferring the dynamics from LME to ENHH, as shown in Figs.~\ref{fig::3} (f) $\sim$ (h). While $\beta$ increases linearly with $\gamma$, it increases nonlinearly with $\eta$, as shown in Figs.~\ref{fig::3} (a) and (i). As for the linear growth, we set $\beta=k\gamma$, then the distribution function in Eq.~(\ref{TanhX}) becomes
\begin{equation}
n(x)=-\frac{1}{2}\tanh\left(k \, \frac{x-(L+1)/2}{1/\gamma} \right) +0.5,
\end{equation}
which implies a scale invariance of steady state in terms of  $1/\gamma$.

\section{Zeno effect versus skin effect}
The NLME of postselected skin effect formally is same as the Hatano-Nelson model in open system. The effective Hamiltonian $H_{eff}(\eta \gamma)$ in Eq.~(\ref{NHT}) leads the non-Hermitian skin effect~\cite{Yao2018,Kunst2018,CHLee2019}, while the environmental effect $\mathcal{L}_{M}$ in Eq.~(\ref{MeaT}) can be understood as continuous measurements, which induces the quantum Zeno effect~\cite{Misra1997}. Considering that both of Zeno effect and skin effect can influence the entangled entropy growth, the TAEE $\S_L(x_C,x_C+\Delta)$ will be a good indicator to reflect the two effects' competition and cooperation. The TAEE is defined as
\begin{equation}\label{TAEE}
\S_L(x_C,x_C+\Delta):=\lim_{N \to + \infty} \frac{1}{N}\sum_{i=1}^N -\Tr \left( \rho^i_{[x_C, x_C+\Delta]} \ln  \rho^i_{[x_C, x_C+\Delta]}   \right),
\end{equation}
where the entropy is an average of $N$ trajectories in a $L$-site chain, and $\rho^i_{[x_C, x_C+\Delta]}$ is the reduced density matrix of the pure state in the $i$-th trajectory on the subinterval $[x_C, x_C+\Delta]$, where $x_C$ is the middle point of the chain and $\Delta$ is the interval length, as shown in Fig.~\ref{fig::2}(b). The time evolution of half-chain entropy $\S_L(x_C,x_C+L/2)$ is shown in Fig.~\ref{fig::2}(c).

As shown in Fig.~\ref{fig::3}(b) and Fig.~\ref{fig::3}(j), the steady-state TAEE exhibits different features in terms of $\gamma$ and $\eta$. It rapidly decreases with increasing $\gamma$, while slowly increases with increasing  $\eta$. The differences come from the combining of Zeno effect and skin effect. When $\gamma$ increases, TAEE is suppressed both by the increasing strength of $\mathcal{L}_{M}$ and the larger nonreciprocity of $H_{eff}(\eta \gamma)$. When $\eta$ increases, the TAEE is governed by the competition between the decrease in Zeno effect and increase in skin effect. The above processes are expressed as
\begin{subequations}
\begin{align}
& \gamma \uparrow\  \Longrightarrow \ \begin{cases}\  \text{Zeno effect} \uparrow \ \Longrightarrow \ \text{Entropy} \downarrow \\ \ \text{Skin effect} \uparrow \ \Longrightarrow \text{Entropy} \downarrow  \end{cases} \\
& \eta \uparrow\  \Longrightarrow \ \begin{cases}\  \text{Zeno effect} \downarrow \ \Longrightarrow \ \text{Entropy} \uparrow \\ \ \text{Skin effect} \uparrow \ \Longrightarrow \text{Entropy} \downarrow  \end{cases}
\end{align}
\end{subequations}
Specifically, the Fig.~\ref{fig::3}(j) shows that in the process of increasing postselection strength, the measurement effect plays a dominant role, where the decreasing Zeno effect overcomes the suppression on entanglement from skin effect and eventually makes the TAEE slowly increase.

\section{Distribution and scaling of entangled entropy}
\begin{figure}[htbp]\centering
\includegraphics[width=8.5cm]{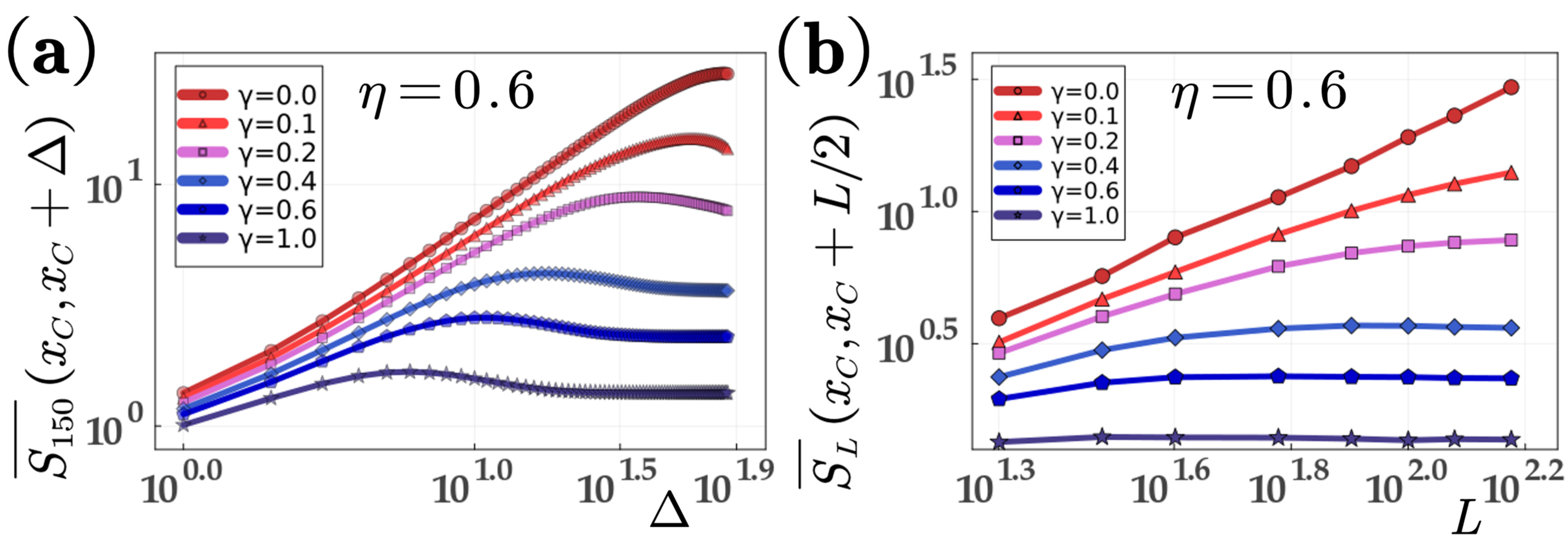}
\caption{ Distribution and scaling of steady-state  entangled entropy. (a) Steady-state entanglement entropy $\S_L(x_C,x_C+\Delta)$ as a function of the bipartition size $\Delta$ in a half-filled $150$-site chain with $\eta=0.6$ and $\gamma$ from $0$ to $1$ in the log-log plot.  (b) Half-chain entanglement entropy $\S_L(x_C,x_C+L/2)$ scaling with the half-filled chain length $L$ in the log-log plot. The entropy has a two-segment scaling behavior: rapid growth in small $L$ and saturation in large $L$. With decreasing $\gamma$, the rapid growth segment becomes longer. Especially in $\gamma=0$, the growth segment increases to the whole chain, which implies a volume law of entropy in the non-dissipative free systems.}
\label{fig::4}
\end{figure}

The postselected skin effect profoundly affects the distribution and scaling behaviors of TAEE by the tanh-fitting steady state, where the particle number only changes rapidly in the middle of the chain. Near the ends of a long chain, the local site is nearly fully occupied ($n_x \approx 1$) or near vacuum ($n_x \approx 0$), which contributes little to entangled entropy. This makes the entropy distribution $\S_L(x_C,x_C+\Delta)$ exhibit a two-segment structure---a nearly algebraic growth with a small $\Delta$ followed by saturation for large $\Delta$, as shown in Fig.~\ref{fig::4}(a).

The steady state's scaling behavior of the half-chain TAEE $\S_L(x_C,x_C+L/2)$ also presents a two-segment structure. When increasing $L$, the population distribution of the whole chain becomes a larger part of the tanh function Eq.~(\ref{TanhX}), as shown in Fig.~\ref{fig::4}(b), and meanwhile the end-site particle number $n_L$ decreases rapidly from $0.5$ and then approaches slowly to $0$. This leads to $\S_L(x_C,x_C+L/2)$ having a nearly algebraic growth in small $L$ and saturation in large $L$, as shown in Fig.~\ref{fig::4}(b).

The above analysis can help us understand the transition from the area law of TAEE with skin effect to the volume law in non-dissipative free systems ($\gamma=0$). Due to $\beta=k\gamma$, decreasing $\gamma$ will make the tanh function change slower, which brings a bigger rapid growth interval. When $\gamma=0$, the whole chain is in the growth segment, giving rise to the volume law of entangled entropy.

\section{Summary and discussion}
In summary, we proposed a NLME to describe the normalized postselection dynamics of Lindblad master equations with loss of quantum jumps. It continuously interpolates between LME and  ENHH by reducing quantum jumps. With the advanced analytical ability of NLME, we answer the question of how it influences the dynamics of LME by losing quantum jumps. We classify the influences into two classes. In trivial class, discarding jumps is completely equivalent to weakening the coupling strength of the environment. While in the nontrivial class, it makes a difference by breaking the balance of LME between the effective non-Hermitain Hamiltonian and quantum jumps. We show a new phenomenon called the postselected skin effect, where the steady state exhibits a special biased particle profile by discarding a few quantum jumps. Moreover, we show the coexistence and competition between Zeno effect and non-Hermitian skin effect in the postselected skin effect by capturing the characteristics of the TAEE.

Our findings offer valuable insights into non-Hermitian physics both in experiment and theory. First, the nontrivial class can reflect non-Hermitian physics through postselection experiments by discarding only a few quantum jumps. We anticipate it offers a new approach to non-Hermitian physics in various experimental settings such as superconducting qubit system~\cite{Naghiloo2019,WChen2021}, cold atomic systems~\cite{JLi2019,WGou2020} and optical systems~\cite{LXiao2020,CWang2023}. Second, NLME can be formally understood a non-Hermitain system embedded in open environment. It gives a reasonable theoretical framework for studying the interplay and competition between the non-Hermitian Hamiltonians and pure dissipations or pure measurements. In this way, we can embed more non-Hermitian problems such as the non-Hermitian topology~\cite{ZGong2018,NOkuma2023} and non-Herminian Anderson localization~\cite{HJiang2019,DWZhang2020} into open systems for future studies.

\section*{Acknowledgements}
We thank Y.-P. Wang for helpful discussions on the Gaussian-state method.
This work is supported by National Key Research and Development Program of China (Grant No. 2023YFA1406704 and 2021YFA1402104), the NSFC under Grants No. 12174436 and No. T2121001, and the Strategic Priority Research Program of Chinese Academy of Sciences under Grant No. XDB33000000.

\appendix
\section{The derivation of the nonlinear Lindblad master equation} \label{AP:1}

Quantum trajectory methods interpret the Lindbald master equation (LME) Eq.~(1) in the main text as a stochastic average over individual trajectories of pure states.  In one trajectory, the system is assumed in $|\phi (t)\rangle$ at time $t$. After $\delta t$, the $L_\mu$ will make the system jump to the state $\N L_\mu |\phi (t)\rangle$ with the probability of $\delta p_\mu=\gamma_\mu \delta t \langle \phi(t)|L^\dag_{\mu}L_{\mu} |\phi(t) \rangle$ or non-unitarily evolve to the state $\N exp(-\frac{1}{2}\gamma_\mu L^\dag_{\mu}L_{\mu}\delta t)|\phi (t)\rangle$ with the probability of $1-\delta p_\mu$. The different jump operators can be regarded as independent measurements by the multiplication rule in probability when $\delta \to 0$. So the state at $t+\delta t$ is
\begin{equation}\label{QTState}
\begin{split}
|\phi(t+\delta t)\rangle= &e^{-iH\delta t} \\
&\N  \left\{ \begin{matrix}
(1) \left\{ \begin{matrix} \delta p_1 : & L_1 \\ 1- \delta p_1 : & \exp(-\frac{1}{2}\gamma_1 L^\dag_{1}L_{1}\delta t)  \end{matrix} \right. \\
\vdots \\
(\mu) \left\{ \begin{matrix} \delta p_\mu : & L_\mu \\ 1- \delta p_\mu : & \exp(-\frac{1}{2}\gamma_\mu L^\dag_{\mu}L_{\mu}\delta t)  \end{matrix} \right. \\
\vdots \\
(M) \left\{ \begin{matrix} \delta p_M : & L_M \\ 1- \delta p_M : & \exp(-\frac{1}{2}\gamma_M L^\dag_{\mu}L_{M}\delta t)  \end{matrix} \right. \\
\end{matrix} \ \ \ \right\} |\phi(t)\rangle.
\end{split}
\end{equation}

Now, we monitor the jump from $L_\mu$ by a detector with efficiency $\eta_\mu$ ($0 \le \eta_\mu \le 1$) and postselect the trajectories where no jumps are detected. Considering the finite detection efficiency, only $(1-\eta_\mu)$ jumps are reserved. This makes the effective possibility of jumping become
\begin{equation}\label{Rjp}
\frac{(1-\eta_\mu)\delta p_\mu}{1-\eta_\mu \delta p_\mu} \approx (1-\eta_\mu)\delta p_\mu,
\end{equation}
and the effective possibility of non-unitary evolution become
\begin{equation}\label{Rnu}
\frac{1-\delta p_\mu}{1-\eta_\mu \delta p_\mu} \approx  1-(1-\eta_\mu) \delta p_\mu.
\end{equation}
Therefore, averaging these trajectories is equivalent to redistributing the probabilities of measurement results by the substitution $\delta p_\mu \to (1-\eta_\mu)\delta p_\mu$. Then the state at $t+\delta t$ is
\begin{equation}\label{PostState}
\begin{split}
& |\phi(t+\delta t)\rangle= e^{-iH\delta t} \\ 
&\N  \left\{ \begin{matrix}
(1) \left\{ \begin{matrix} (1-\eta_1)\delta p_1 : & L_1 \\ 1- (1-\eta_1)\delta p_1 : & \exp(-\frac{1}{2}\gamma_1 L^\dag_{1}L_{1}\delta t)  \end{matrix} \right. \\
\vdots \\
(\mu) \left\{ \begin{matrix} (1-\eta_\mu)\delta p_\mu : & L_\mu \\ 1- (1-\eta_\mu)\delta p_\mu : & \exp(-\frac{1}{2}\gamma_\mu L^\dag_{\mu}L_{\mu}\delta t)  \end{matrix} \right. \\
\vdots \\
(M) \left\{ \begin{matrix} (1-\eta_M)\delta p_M : & L_M \\ 1- (1-\eta_M)\delta p_M : & \exp(-\frac{1}{2}\gamma_M L^\dag_{\mu}L_{M}\delta t)  \end{matrix} \right. \\
\end{matrix} \ \ \ \right\} |\phi(t)\rangle.
\end{split}
\end{equation}
Due to that $\delta p_\mu$ is the first order small quantity of $\delta t$, in the above stochastic evolution, the most possible event is non-unitary evolution for all $\mu$ and the second most probable event is only one jump with $M-1$ non-unitary evolutions.

Making the first-order approximation of $\delta t$, the maximum probability state $|\phi_{eff}\rangle$ and its probability $P_{Max}$ in Eq.~(\ref{PostState}) are
\begin{equation}
P_{Max} = \prod_\mu [1-(1-\eta_\mu)\delta p_\mu] \approx 1-\sum_\mu (1-\eta_\mu) \delta p_\mu,
\end{equation}
\begin{equation}
\begin{split}
|\phi_{eff}\rangle &= e^{-iH\delta t} \N \left( \prod_\mu \exp(-\frac{1}{2}\gamma_\mu L^\dag_{\mu}L_{\mu}\delta t) \right) |\phi (t)\rangle \\
                   &\approx \N \exp(-iH_{eff}\delta t) |\phi (t)\rangle \\
                   &= \frac{1}{\sqrt{1-\sum_\mu \delta p_\mu}} \exp(-iH_{eff}\delta t) |\phi (t)\rangle, \\
\end{split}
\end{equation}
where $H_{eff}(\gamma)=H-\frac{i}{2}\sum_\mu \gamma_\mu L_{\mu}^{\dag} L_{\mu}$. The second highest probability comes from jumping only once in $M$ measurements. For example, only jumping at $L_\mu$ will output the state $|\phi_\mu\rangle$ and its probability $P_\mu$:

\begin{equation}
P_\mu = (1-\eta_\mu)\delta p_\mu \prod_{j\ne \mu} [1-(1-\eta_j)\delta p_j] \approx (1-\eta_\mu) \delta p_\mu,
\end{equation}
\begin{equation}
\begin{split}
|\phi_\mu\rangle &= e^{-iH\delta t} \N \left( \prod_{j\ne\mu} \exp(-\frac{1}{2}\gamma_j L^\dag_j L_j \delta t) \right) L_\mu |\phi (t)\rangle \\
                 &\approx \N  \exp(-iH \delta t) L_\mu |\phi (t)\rangle \\
                 &= \frac{1}{\sqrt{ <L_{\mu}^{\dag} L_{\mu}> }} \exp(-iH \delta t) L_\mu |\phi (t)\rangle. \\
\end{split}
\end{equation}

Including the above two kinds of states and making the first-order approximation of $\delta t$, the nonlinear Lindblad master equation (NLME) is obtained by
\begin{widetext}
\begin{equation}\label{DNLME}
\begin{split}
\frac{d}{dt} \rho &= \lim_{\delta t \to 0} \frac{1}{\delta t} (|\phi(t+\delta t)\rangle \langle \phi(t+\delta t) | - |\phi(t)\rangle \langle \phi(t) |) \\
&= \lim_{\delta t \to 0} \frac{1}{\delta t} \left( P_{Max}  |\phi_{eff}\rangle \langle \phi_{eff} | + \left(\sum_\mu P_\mu |\phi_\mu \rangle \langle \phi_\mu| \right) - |\phi(t)\rangle \langle \phi(t) | \right) \\
&= \lim_{\delta t \to 0} \frac{1}{\delta t} \left( \frac{1-\sum_\mu (1-\eta_\mu) \delta p_\mu}{1-\sum_\mu \delta p_\mu} e^{-iH_{eff}\delta t} \rho e^{iH^\dag_{eff}\delta t} + \sum_\mu \left( \frac{ (1-\eta_\mu) \delta p_\mu }{<L_{\mu}^{\dag} L_{\mu}>} e^{-iH \delta t} L_\mu \rho L^\dag_\mu e^{iH \delta t} \right) - \rho \right) \\
&= \lim_{\delta t \to 0} \frac{1}{\delta t} \left( (1+\sum_\mu \eta_\mu \delta p_\mu) [\rho-i\delta t (H_{eff}\rho-\rho H^\dag_{eff})] + \sum_\mu \left( (1-\eta_\mu)\gamma_\mu \delta t L_\mu \rho L^\dag_\mu  \right) -\rho \right) \\
&= -i[H,\rho] + \sum_\mu \gamma_{\mu} \left( -\frac{1}{2} \{ L_{\mu}^{\dag} L_{\mu}, \rho \} + (1-\eta_\mu) L_{\mu} \rho L_{\mu}^{\dag} + \eta_\mu  <L_{\mu}^{\dag} L_{\mu}> \rho \right).\\
\end{split}
\end{equation}
\end{widetext}

\section{Vectorization treatment of the two-level atom} \label{AP:2}

The Eq.~(\ref{AtomE}) in the main text describes the postselection dynamics of the two-level atom with monitoring its spontaneous emission. We can deal with it by vectorization method~\cite{Hugo2023}. First the density matrix $\rho$ is mapped to the density vector $|\rho\rangle$:
\begin{equation}
\rho= \left(\begin{matrix}  \rho_{ee} & \rho_{eg} \\ \rho_{ge} & \rho_{gg} \end{matrix} \right) \ \ \to \ \ |\rho\rangle= (\rho_{ee}, \  \rho_{eg}, \ \rho_{ge}, \  \rho_{gg})^T.
\end{equation}
Second, the NLME is mapped as
\begin{equation}
\frac{d}{dt}\,|\rho\rangle\,=\,\RL_\eta \, |\rho\rangle,
\end{equation}
where
\begin{equation}
\begin{split}
\RL_\eta= & -i\RH \otimes \RI + i \RI \otimes \RH^{\RT} -\frac{\gamma}{2} (\RL^\dag \RL \otimes \RI + \RI \otimes \RL^{\RT} \RL^* ) \\
          &+ (1-\eta)\gamma \RL \otimes \RL^* + \eta \gamma <\RL^\dag \RL>,\\
\end{split}
\end{equation}
where
\begin{equation}
\RH=\left( \begin{matrix} 0 & J \\ J & 0  \end{matrix} \right), \ \ \RL=\left( \begin{matrix} 0 & 0 \\ 1 & 0  \end{matrix} \right), \ \ \RI=\left( \begin{matrix} 1 & 0 \\ 0 & 1  \end{matrix} \right).
\end{equation}
Then we get the vectorized NLME:
\begin{equation}
\frac{d}{dt}\left(\begin{matrix} \rho_{ee} \\ \rho_{eg} \\ \rho_{ge} \\ \rho_{gg} \end{matrix}  \right) =
\left(\begin{matrix}
-iJ(\rho_{ge}-\rho_{eg}) +\eta \gamma  \rho_{ee}^2 - \gamma \rho_{ee} \\
-iJ(\rho_{gg}-\rho_{ee}) + \eta \gamma \rho_{eg} \rho_{ee} -0.5\gamma\rho_{eg} \\
iJ(\rho_{gg}-\rho_{ee}) + \eta \gamma \rho_{ge} \rho_{ee} -0.5\gamma\rho_{ge} \\
iJ(\rho_{ge}-\rho_{eg}) + \eta \gamma \rho_{gg} \rho_{ee} + (1-\eta)\gamma\rho_{ee}.
\end{matrix}  \right)
\end{equation}
From the above equation we can obtain the evolution of $\rho$ by the Runge-Kutta methods.

\section{Monte Carlo wave-function method for NLME} \label{AP:3}

The Monte Carlo wave-function method also called the quantum trajectory method, was originally proposed as a numerical method for simulating dissipative dynamics in the early 1990s~\cite{Dalibard1992,Dum1992,Gardiner1992,Klaus1993,Plenio1998}. This method rewrites the LME as a stochastic average over individual trajectories. Compared with the density matrix in LME, the trajectory, which is the time-dependent wave function, significantly saves memory resources.

Applying the Monte Carlo wave-function method on the NLME in Eq.~(\ref{NLME})  in the main text, the postselection processes will influence the probability distribution of random quantum jumps and non-unitary evolution in trajectories. For a single trajectory, we assume the wave function of the system reaches the state $|\phi (t)\rangle$ at time $t$. In the following time of $\delta t$, the system Hamiltonian will bring a unitary evolution by $e^{-iH\delta t}$. For every dissipation term, for example, the $L_\mu$ will make the system jump to the state $\N L_\mu |\phi (t)\rangle$ with the possibility $\delta p_\mu=\gamma_\mu \delta t \langle \phi(t)|L^\dag_{\mu}L_{\mu} |\phi(t) \rangle$ or non-unitarily evolve to the state $\N exp(-\frac{1}{2}\gamma_\mu L^\dag_{\mu}L_{\mu}\delta t)|\phi (t)\rangle$. Due to the postselection with efficiency $\eta_\mu$, the jumping data with the possibility of $\eta_\mu \delta p_\mu$ is discarded. Therefore, after normalizing all trajectories, the possibility of the jumping is updated to $\frac{(1-\eta_\mu)\delta p_\mu}{1-\eta_\mu \delta p_\mu} \approx (1-\eta_\mu)\delta p_\mu$ and the possibility of the non-unitary evolution is updated to $\frac{1-\delta p_\mu}{1-\eta_\mu \delta p_\mu} \approx  1-(1-\eta_\mu) \delta p_\mu $. The above processes are expressed as the Eq.~(\ref{tadt}) in the main text.

To calculate the average value of the steady state, we simulate a long enough time for the system to reach the saturation value by the Monte Carlo method. Then we average over the data at the end of the period, for example, the TAEE of steady state is an average from $t=270$ to $t=300$ in Fig.~\ref{fig::2}(c) in the main text.

\section{Numerically confirm the validity of NLME} \label{AP:4}

There are three methods to simulate the postselection dynamics in this work. We compare the numerical results from three methods to confirm the validity of NLME.

\begin{figure}[htbp]\centering
\includegraphics[width=8.5cm]{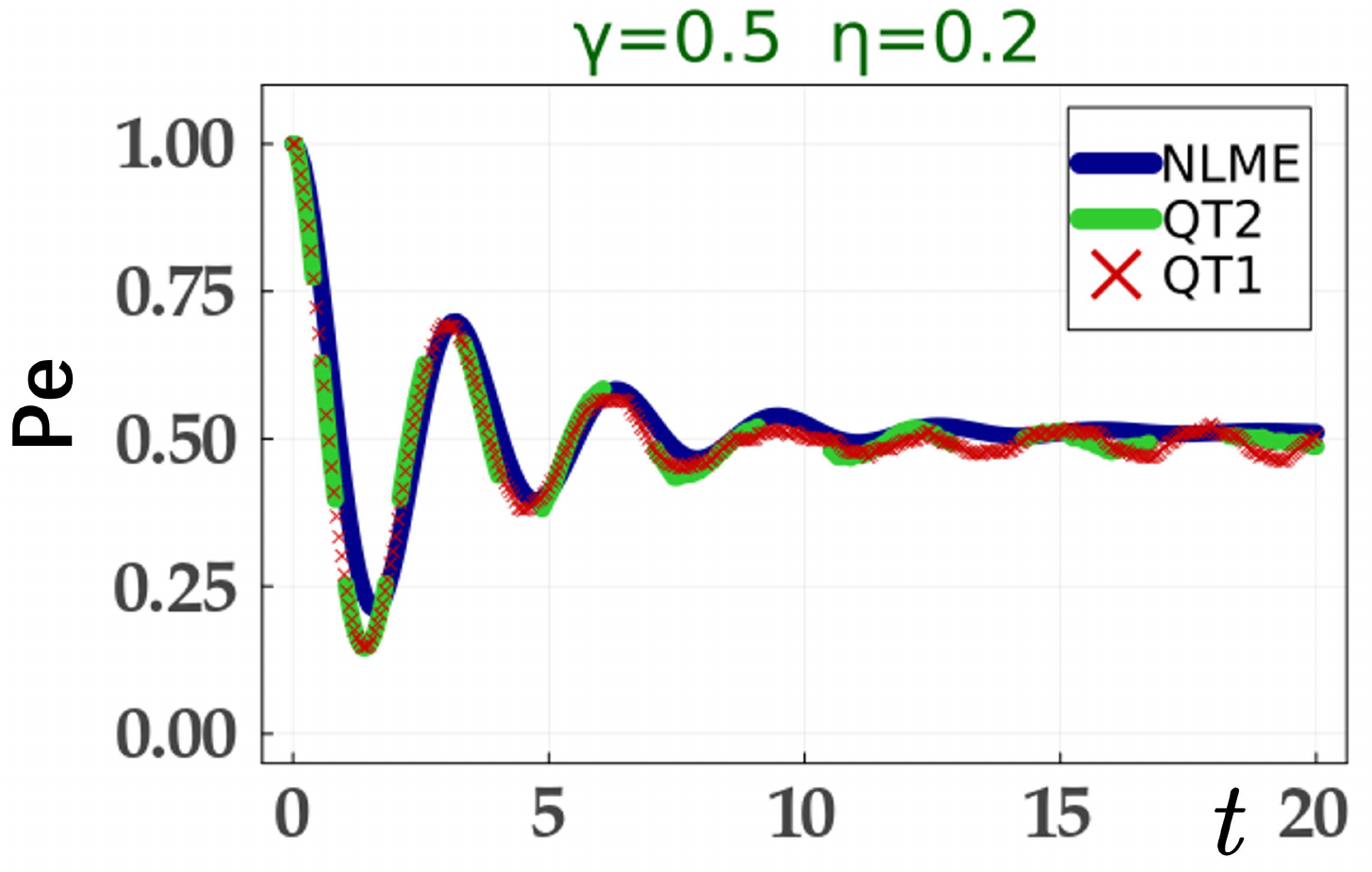}
\caption{The simulation of excited state probability from three methods. The system is described by Eq~(\ref{AtomE}) in the main text with $\gamma=0.5$ and $\eta=0.2$. The atom is initialized in the excited state $|e\rangle$. The excited state probability is simulated by the methods ``QT1''(red cross), ``QT2''(green dashed line) and ``NLME''(blue solid line). }
\label{fig::5}
\end{figure}

(1) According to the proposal of experimental implementation as shown in Fig.~\ref{fig::1}(a) in the main text, the postselection processes need to monitor dissipations and collect the experimental data without detectors clicking. The most faithful simulation is based on the original quantum trajectory method with postselecting the trajectories, which we mark as the method ``QT1''. The process is as follows. First, We simulate the system evolution by Eq.~(\ref{QTState}). For every decay channel in every time step $\delta t$, for example, $L_\mu$ in time $t \sim t+\delta t$, we use a random number $r_1 \in [0,1)$ to choose the evolution processes. If $r_1 < \delta p_\mu$, the quantum jump occurs and evolution is $|\phi (t+\delta t)\rangle = \N L_\mu |\phi (t)\rangle$. If $r_1 \ge \delta p_\mu$, the non-unitary evolution occurs, i.e., $|\phi (t+\delta t)\rangle = \N exp(-\frac{1}{2}\gamma_\mu L^\dag_{\mu}L_{\mu}\delta t)|\phi (t)\rangle$. Second, if quantum jumps occur, we use a second random number $r_2 \in [0,1)$ to decide whether the jump will be detected. If $r_2 < \eta_\mu$, jumps is detected and we discard this trajectory. Only the trajectories with no jumps being detected are saved. Third, we use the saved trajectories to calculate the observation average values.

(2) Discarding quantum jumps from postselection is equivalent to increasing (decreasing) the probability of the processes of non-unitary evolution (quantum jumps). The second simulation method is just adjusting the probability accroding to the postselection effciency $\eta_\mu$ in the original quantum trajectory method as shown in by Eq.~(\ref{Rjp}) and Eq.~(\ref{Rnu}). Therefore, we can calculate the postselection dynamics by the ensemble of trajectories produced by Eq.~(\ref{PostState}) without further postselecting the trajectories. We mark the second method as ``QT2''.

(3) We use the method "QT2" to derive the nonlinear Lindblad master equation (NLME) in Eq.~(\ref{DNLME}). However, the NLME as the ordinary differential equation can be directly simulated by some numerical methods such as Runge-Kutta methods. We mark the simulation directly from the NLME as the method ``NLME''.

We simulate the postselection dynamics of the two-level atom with monitoring its spontaneous emission in Eq~(\ref{AtomE}) in the main text by the above three methods. The three curves of time evolution of excited state probability are almost the same as shown in Fig.~\ref{fig::5}. This confirms the validity of NLME.

\section{The Gaussian state method for free fermions} \label{AP:5}
The model of the postselected skin effect is a free fermionic system. In our Monte Carlo simulation, it starts at a product state and keeps the Gaussian state in every quantum trajectory. This process can be effectively simulated by the Gaussian state method~\cite{Kawabata2023,Wang2022,KLi2023}. Here we show how to calculate this dynamics.

Consider $N$ fermions in a $L$ state chain ($N \le L$). Define the vector creation and annihilation operators corresponding to a non-normalized column vector $\bv \in \mathbb{C}^L$ as
\begin{subequations}
\begin{align}
& \hdd_{\bv}=(c^\dag_1, c^\dag_2, \cdots c^\dag_L) \cdot \bv = \sum_{i=1}^{L} v_i c^\dag_i, \\
& \hd_{\bv}= \bv^\dag  \cdot (c_1, c_2, \cdots c_L)^\RT = \sum_{i=1}^{L} v_i^* c_i.
\end{align}
\end{subequations}
A $N$-particle product state is defined as $|V\rangle = \prod_{n=1}^N \hdd_{\bv_n} |0\rangle$, where $|0\rangle$ is the vacuum state and $V$ is a $L \times N$ matrix defined as $V=\{\bv_1, \bv_2, \cdots \bv_N \}$. $|V\rangle$ can be normalized by the QR decomposition $V=QR$, where $Q=\{\bq_1, \bq_2, \cdots \bq_N \}$ is a $L \times N$ matrix satisfying $Q^\dag Q = \RI$ and $R$ is a $N \times N$ upper triangular matrix. Due to Pauli exclusion principle, if $V$ is linearly dependent ($rank(R)<N$), then $\N |V\rangle=0$.
If $V$ is linearly independent ($rank(R)=N$),
\begin{equation}
\N |V \rangle = \N  \prod_{n=1}^N (\sum_{j=1}^n R_{jn} \hdd_{\bq_j})|0\rangle = \prod_{n=1}^N \hdd_{\bq_j} |0\rangle =|Q\rangle,
\end{equation}
where $\hdd_{\bq_i}$ and $\hd_{\bq_j}$ satisfy fermionic anticommutation relations.

The real space correlation matrix $C$ is calculated as $C=Q^* Q^\RT$, where $C_{ij}=\langle Q| c^\dag_i c_j |Q \rangle$ and the bipartite entanglement entropy $S_L(a,b)$ for subinterval $[a,b]$ in $L$ chain equals
\begin{equation}
S_L(a,b)=-\sum_i \left[ \lambda_i ln \lambda_i  + (1-\lambda_i) ln (1-\lambda_i)     \right],
\end{equation}
where $\lambda_i$ is the eigenvalue of the truncated $C$ in the interval $[a,b] \times [a,b]$.

In every $\delta t$ of the simulation, there are three kinds of operators: the system unitary evolution $e^{-iH\delta t}$, the non-unitary evolution $\N exp(-\frac{1}{2}\gamma_l L^\dag_l L_l \delta t)$, and the quantum jump $\N d^\dag_l d_l$. The first two operators can be uniformly expressed as $e^{\hat{J}_M} $ with a quadratic form
\begin{equation}
\hat{J}_M = (c^\dag_1, c^\dag_2, \cdots c^\dag_L) M (c_1, c_2, \cdots c_L)^\RT,
\end{equation}
where $M$ is a $L \times L$ matrix. By the relation
\begin{equation}
e^{\hat{J}_M} c^\dag_j e^{-\hat{J}_M} = \sum_m c^\dag_m (e^M)_{mj},
\end{equation}
we get
\begin{equation}
\begin{split}
\N e^{\hat{J}_M} |V\rangle &= \N \prod_{n=1}^N \left( \sum_{j=1}^L  e^{\hat{J}_M} c^\dag_j e^{-\hat{J}_M} V_{jn}  \right) |0\rangle \\
                          &= \N \prod_{n=1}^N \left( \sum_{m=1}^L c^\dag_m (e^M V)_{mn} \right) |0\rangle = \N | e^M V \rangle. \\
\end{split}
\end{equation}
The jump operators can be decomposed into the vector operator of $\hdd_{\ba}$ or $\hd_{\ba}$ on product states. The $\N \hdd_{\ba} |V\rangle$ equals adding basis $\ba$ into space $V$, i.e.
\begin{equation}
\N \hdd_{\ba} |V\rangle = \N | \ba \otimes V \rangle := \N | \{\ba, \bv_1, \bv_2 \cdots \bv_N \} \rangle.
\end{equation}
By the relation
\begin{equation}
\hd_{\ba}  \hdd_{\bv} = (\sum_i \alpha^*_i c_i) (\sum_j v_j c^\dag_j) = \ba^\dag \cdot \bv - \hdd_{\bv} \hd_{\ba},
\end{equation}
we have
\begin{equation}
\hd_{\ba} |V\rangle = \sum_{l=1}^N (-1)^{l-1} (\ba^\dag \cdot \bv_{l} ) \prod_{n \ne l} \hdd_{\bv_n} |0\rangle,
\end{equation}
and $\langle V| \hdd_{\ba} \hd_{\ba} |V \rangle =\sum_{l=1}^N (\ba^\dag \cdot \bv_{l} )^2 $. The geometric meaning of $\N \hd_{\ba} |V\rangle$ is finding the orthogonal complement of $\ba^{//}$ in $V$, where $\ba^{//}$ is the projection of $\ba$ onto $V$. This process can be calculated by
\begin{equation}
\N \hd_{\ba} |V\rangle = \N \  | \ {\rm ker} \left( (\ba \otimes {\rm ker} (V^\dag ))^\dag  \right) \  \rangle,
\end{equation}
where ``${\rm ker}()$'' is to find the null space of a matrix.



\end{document}